\documentstyle[11pt]{article}

\begin{document}

\baselineskip = 18pt plus 1pt minus 1pt

\title{New metastable Charmonium and the $\psi '$ anomaly at CDF}

\author{FE Close\\
\makebox[5cm]{}
\\
{\em Rutherford Appleton Laboratory,}\\
{\em Chilton Didcot, Oxon OX11 0QX, Great Britain.}
\\
FEC@v2.rl.ac.uk
\\
\\
RAL--94--093 \\
\\
\\
}
\vspace{2cm}
\date{ August 1994}

\maketitle

\begin{abstract}
\noindent Production of metastable charmonium states more massive than the
$\psi(3685)$ is expected in models and could be the source of up to $50\%$
of the $\psi$ observed at large $p_T$ at the Tevatron.Narrow $2^{-+},2^{--}
c\bar{c}$ are predicted at around 3.8GeV and radially excited
$2^3P_{1,2}$ may also have suppressed hadronic widths making these states
potentially extra sources of $\psi(3685)$.
Colour octet components are believed to dominate $\psi$ production at the
Tevatron which suggests that hybrid charmonium production may also be
prominent. Estimates of hybrid production rates and branching ratios into
charmonium suggest that metastable hybrids with mass $\approx$ 4GeV may
play an important role in generating the observed $\psi(3685),\psi(3095)$
\end{abstract}
\pagestyle{empty}

\newpage
\pagenumbering{arabic}
\pagestyle{plain}

\section{Introduction}

The largest discrepancy between the predictions of the Standard Model
and experiment may be the recent measurements of charmonium
production at large transverse
momentum in $p\bar{p}$ collisions at the Tevatron\cite{CDF}.
The clearest evidence for an anomaly comes from the
CDF report of a $\psi '$ production rate that
 is a factor of about 30 larger than theoretical expectation, too
large a discrepancy to be accomodated by tinkering with parameters in the
theory
and suggestive of a source additional to those so far included in the
calculations.
Production of the $\psi$ also appears to be enhanced relative to expectations
 by a factor of 2 or 3 \cite{CDF};
 though smaller than the $\psi '$ enhancement
in ratio, this is nonetheless significant and a comparable enhancement in
absolute magnitude.

In this letter I bring together recent developments in, at first sight,
rather unrelated areas of theory and experiment and show how they may provide
a consistent picture including a possible explanation of the charmonium
anomaly. In summary; there are reasons to suspect that
metastable charmonium states exist more massive than the $\psi '$, that
their production rates in hadronic interactions are
comparable to the $\chi$ states and that their decays
could feed the lower mass signals leading to a significant enhancement
over the present theoretical expectations. Three main candidates occur:

(i) conventional charmonium states $2^{-+},2^{--}$, predicted at 3.81 to
3.85GeV
\cite{godfrey,Quigg} and thereby metastable (parity forbids the $D\bar{D}$
decay),

(ii) the radial excitation  $2^3P_1(\approx 3.9GeV)$ whose decay
$\rightarrow DD^*$ is near threshold where radial wavefunctions can
suppress widths\cite{ley,koko,page} and its partner $2^3P_2(3.9-4.0GeV)$ whose
decays into $D\bar{D}, D\bar{D^*}$ are similarly suppressed by $D$-wave phase
space and dynamical effects,

(iii)hybrid charmonium states where the gluonic degrees of freedom are
dynamically excited in the presence of charmed quarks \cite{has}-\cite{slear}.
 Some of these
states are
predicted to be metastable if their mass is below $DD^{**}$ threshold
($\approx 4.3GeV$)\cite{IP,Lipkin,Iddir}; recent works \cite{IP,mich,BCS,slear}
have reinforced the predictions\cite{has,BCD,slear} that
this mass region is the most likely for the manifestation of hybrid
charmonium states.

These three scenarios have competing claims. The first is conservative and
almost certain to be present at some level: the problem is that
the decays into the $\psi$ and $\chi$ charmonia are suppressed by
phase space and/or wavefunction orthogonality (however,the radiative decay of
$2^{-+}
\rightarrow \psi(3685) + \gamma$ may become detectable if the $\psi(3685)$
contains significant $^3D_1$ or other non trivial components in its
wavefunction). The radiative transitions
 $2^3P_{1,2} \rightarrow 2^3S_1 + \gamma$ could be significant
 sources of $\psi(3685)$ if the hadronic width of either $2^3P_2$ or $2^3P_1$
 is suppressed, as suggested in some models
\cite{koko,page,page1}.The hybrid excitation on the other hand,
is more radical and in consequence more
interesting. If such states exist in Nature, then production by hard gluons
in a hadronic environment favouring heavy flavours, such as at the Tevatron,
 may be rather natural.
The pathways to feed $\psi$ and $\chi$ charmonium states are plentiful
and some enhancement of $\psi$ states may be expected. In any event,
the definitive test of all of these scenarios
is to form invariant mass distributions of $\psi$ $\chi$
and $\psi '$ states in combination with $\gamma, n\pi, \eta$ in order to
isolate any source(s) more massive than $\psi(3685)$.

Our initial point of departure will be the recent series of papers
 \cite{Bra1,Bra2,Bra,Li}
on the production and decay of heavy quarkonia which have, {\it inter alia},
stressed the important role that colour-octet components can play, especially
in
the case of $P$-wave (and, we conjecture, $D$-wave) quarkonia. The essential
physical ideas are as follows.

Any meson is a superposition of many components

\begin{equation}
M = \psi_{Q\bar{Q}} + O(\vec{v})\psi_{Q\bar{Q}g} + \ldots
\end{equation}
in some of which ($ e.g. \psi_{Q\bar{Q}g}$) the heavy quark and antiquark are
in
a colour-octet state. The probability of the $Q\bar{Q}g$ state is of order
$\vec{v}^2$ (heavy quarks don't easily radiate a gluon;$\vec{v}^2 \approx
\frac{1}{4}$ for charmonium ) and so such components
can be neglected for many applications. However, as first pointed out in ref
\cite{Bra2}, these components play an enhanced role in the decay and
production of
P states where the $Q\bar{Q}g$ contribution to the quarkonium annihilation
competes with, and in some cases even dominates, that coming from
the $Q\bar{Q}$.
The essential physics is that the $Q\bar{Q}$ contribution to annihilations of
P-wave
quarkonium is suppressed by $v^2$ owing to the angular-momentum barrier which
pushes
the quarks apart; in the $Q\bar{Q}g$ component the $Q\bar{Q}$ are in S-wave
for which there is no angular momentum suppression though the probability to
find this component in the wavefunction is only at $O(v^2)$ and hence the
$Q\bar{Q}$ and $Q\bar{Q}g$ contributions are comparable in production or
decays.

These ideas have been extended and applied to charmonium production
in $B$-decays \cite{Bra1} and to the production of
both $S$-wave and $P$-wave charmonia by gluon fragmentation\cite{Bra}. As
a result of
the colour-octet mechanism, the production of $\chi$ states is predicted to
play a prominent role, in particular their subsequent decay $\chi \rightarrow
\psi + \gamma$
is predicted to dominate over direct production of $\psi$.
There is evidence to support this latter prediction and rather general support
for the prediction that a major component of heavy quarkonia production at
high $p_{T} \geq 6GeV$
is fragmentation (namely the production of high $P_T$ parton, in particular
a gluon, followed by splitting into a quarkonium
state)\cite{Roy,grec,bra3,UA1}.

Even though this important colour octet contribution helps to explain
in part the
large $J/\psi$ production recently observed at the Tevatron,
it still appears to fall short of
explaining the full strength of $\psi$ production, both there
\cite{Roy,grec,bra3} and also
in $B$-decay \cite{Bra1}. The prediction\cite{Bra1} in the latter case is that
$BR(B \rightarrow \psi + \ldots) = 0.23\%$ to be compared with $(0.74 \pm
0.21)\%$
(where the cascade from $\chi$ states has been accounted for and we
have used the new PDG values \cite{PDG94})

Production of one or more additional metastable charmonium state(s) with
masses greater than 3685MeV and with significant branching ratios into
the conventional charmonia could be responsible for the enhanced signals
in the above set of experiments.

\section{Colour Octet States}

Independent of the above, there has been considerable interest in the possible
existence of hybrid states \cite{has,BCD,IP} where the gluonic degrees of
freedom are
excited in the presence of the quarks. Early extensive studies \cite{BCD}
modelled these states
as $Q\bar{Q}g$ where the quarks were in an overall colour octet state;
subsequent developments
have included lattice simulations \cite{mich} and
modelling in terms of flux tubes connecting the quarks \cite{IP}.
 In these latter the conventional mesons arise when the flux tube is in
its ground state and new hybrid states arise when the flux tube is excited.
 In
the case of an unexcited flux tube,
the colour can be factored onto the $Q\bar{Q}$
such that they are colour singlet; when the colour flux tube is excited such
a factorisation is non trivial and overlap with the colour octet simulations
of refs \cite{has,BCD} may arise but a detailed relationship is still unclear.

In contrast to the conventional quarkonium (eq 1), the hybrid states may be
considered as \cite{BCD,Li}

\begin{equation}
H =  \psi_{Q\bar{Q}g} + O(\vec{v})\psi_{Q\bar{Q}} + \ldots
\end{equation}
If for the ground state the $(Q\bar{Q})_8$ are in S-wave
\cite{BCD}, the production of the hybrid state
can be dominated by the colour octet mechanisms of refs \cite{Bra1,Bra2,Bra} in
{\it leading order}.

The overall quantum numbers of the hybrids depend on whether the gluon
coupling is chromo-electric or magnetic\cite{BCD}, forming respectively
positive and negative parity families, $(0,1,2)^{++},1^{+-}$
or $(0,1,2)^{-+},1^{--}$. In particular, among the chromomagnetic
configurations we note the state with $J^{PC}=2^{-+}$
which would be $D-$wave in colour singlet quarkonium, and hence highly
suppressed in short distance production ($O(v^4)$), but produced in
leading order in the octet configuration,

\begin{equation}
D(g \rightarrow Q\bar{Q} 2^{-+}) \approx R''_d(0)^2 \approx v^4 \times R_s(0)^2
\end{equation}
\begin{equation}
D(g \rightarrow Q\bar{Q}g J^{-+}) \approx (2J+1)R_{8,s}^2
\end{equation}
where $R_{8,s}^2$ is the non-perturbative probability for $Q\bar{Q}$
in the S-wave colour octet configuration to fragment into a colour singlet
bound state and is expected, following ref \cite{Bra1,Bra2,Bra},
to be proportional to the probability of finding the $Q\bar{Q}g$
octet component in the hadron wavefunction. Thus if the $Q\bar{Q}g$
``octet modelling" of hybrids\cite{has,BCD,Li} is a guide,
we would expect that any $2^{-+}$ charmonium production at the Tevatron will
favour the hybrid configuration.

In the case of chromoelectric gluon couplings, which lead to the same overall
quantum numbers as those of P-wave charmonium,ref \cite{Bra} finds for the
gluon fragmentation probabilities

\begin{equation}
P_{g \rightarrow \chi_J} \approx (2J+1)\frac{\pi \alpha_s H_8}{24 m_Q}
- R_J \frac{\alpha_s^2 H_1}{108m_Q}
\end{equation}
 where $R_{0,1,2} = 5,4,16$ respectively and $H_1 \approx 15MeV$, $H_8 \approx
3MeV$. If $\alpha_s \approx \frac{1}{4}$, this implies
fragmentation probabilities
of $(0.4,1.8,2.4)10^{-4}$ for $\chi_{0,1,2}$ where in each case the octet
piece dominates. Multiplying the fragmentation probabilities by the
appropriate radiative branching ratios for $\chi_J \rightarrow \psi +
\gamma$ of $(0.7,27,14)10^{-2}$ yields the
probability of $J/\psi$ in a gluon jet of approximately $8 \times 10^{-5}$,
over an order of magnitude larger than the probability $3 \times 10^{-6}$
for direct fragmentation of a gluon into $J/\psi$\cite{Bra}.

 Insofar as the empirical value $H_8 \approx
3MeV$ subsumes the octet probability $O(v^2 \approx \frac{1}{4})$,
we may estimate the probabilities
for fragmentation into the hybrid counterparts by rescaling the $H_1,H_8$
accordingly

\begin{equation}
H_8^{hybrid} \approx 12MeV
\end{equation}
\begin{equation}
H_1^{hybrid} \approx 3.75MeV
\end{equation}
in which case the fragmentation probabilities for the hybrids are

\begin{equation}
P_{g \rightarrow \chi_g}(0^{++},1^{++},2^{++}) = (2,8,13) \times 10^{-4}
\end{equation}
These give a total probability $O(10^{-3})$ which would lead to the
required enhancement of
the $J/\psi$ signal if $B.R.(\psi + \ldots) \approx 4 \times 10^{-2}$.

A problem is that models put these {\it positive} parity hybrid
states significantly above the charmed
hadron pair threshold with consequent strong decay widths $O(100)MeV$.
Therefore we find do not anticipate a large ($O(10^{-2})$) branching
ratio into charmonium unless there is strong mixing with the nearby radial
charmonium states in particular $2^3P_{1,2}(3.9-4.0GeV)$ (see section 3)

By contrast {\it negative parity} hybrids could be a copious source of
the $\psi$ states due to two fortunate circumstances common in models.
First, models based on constituent
gluons\cite{has,BCD},
flux tubes\cite{IP,BCS}, quenched heavy quark lattice QCD\cite{mich}
and QCD sum rules\cite{Narison} all anticipate negative parity hybrid
charmonium
states, including $(0,1,2)^{-+}, 1^{--}$,  typically in the range
$4.2 \pm 0.2GeV$. In addition the negative parity decays into $D\bar{D},
D^*\bar{D}$ are suppressed in both constituent gluon models\cite{Lipkin,Iddir}
and in flux tube models\cite{IP} such that the effective strong decay threshold
is at $DD^{**} \approx 4.2-4.3GeV$ (an estimate of their widths leads
typically to $O(1MeV)$\cite{page}). Thus they offer {\it a priori} the
possibility of radiative branching ratios into $\chi,\psi$ of $O(10^{-3})$
or via hadronic modes at the $10^{-2}$ level if the gluonic tumbledown
transitions $\psi(2S) \rightarrow \psi(1S) \geq 50\%$ and predictions of
similar
rates for $1D$-states\cite{yan} are any guide.

Production of these states in $\gamma \gamma$ processes has been discussed
in\cite{Li} but I know of no detailed
calculations analogous to\cite{Bra} of gluonic
fragmentation, factoring out chromomagnetic and electric gluon modes.
 Ref\cite{Li} found that in $\gamma \gamma$ decays the colour singlet and octet
components compete (for the hybrid decay)
and cancel in perturbation theory; in the hadronic decays
this cancellation does not occur and the colour octet component is dominant.
Optimistically one could take over the results of
equation(8) to the negative parity case in which case one finds a significant
possibility of $\psi$ production competing with the existing signal (or
for the $\psi(3685)$ a dominant effect). However, we suspect that a more
realistic calculation will find that the results for the negative parity
states will effectively be reduced by $O(v^2)$ in probability relative
to those for the
analogous positive parity states, (the $g^*g \rightarrow Q\bar{Q}g$
transition with magnetic $J^P=1^+$, involves an effective $P-$wave ($^3P_1$)
coupling
relative to the $^3S_1$ picture and so we will form our more conservative
estimate by reducing the octet contribution in equation(5) by $O(v^2) \approx
\frac{1}{4}$).

The singlet contribution in equation(5) also needs to be reconsidered. For
the $0^{-+}$ hybrid, the colour singlet component is $^1S_0$ and so the
effective contribution

\begin{equation}
H_1(0^{-+}) \approx \frac{3}{2\pi} \frac{R_s(0)^2}{m_Q^4} \times 0(v^2)
\approx 27MeV
\end{equation}
For the case of the exotic $1^{-+}$ and the $2^{-+}$ we need only include
the octet piece in our estimates (see eq(15) {\it et seq} below).
 So these lead to our ``optimistic" and ``conservative" ($0(v^2)$ suppression)
estimates for the production probabilities for the negative
parity hybrids in the $Q\bar{Q}g$ picture

\begin{equation}
3 \times 10^{-4} \leq P(2^{-+}) \leq 13 \times 10^{-4}
\end{equation}
\begin{equation}
2 \times 10^{-4} \leq P(1^{-+}) \leq 4 \times 10^{-4}
\end{equation}
\begin{equation}
10^{-5} \leq P(0^{-+}) \leq 1 \times 10^{-4}
\end{equation}
While these follow rather naturally in the $Q\bar{Q}g$ picture they are
probably only an order of magnitude guide for the flux tube model of hybrids.
More detailed study is warranted in the latter case as the current
estimates are encouraging: in both models the hadronic widths are of order
$1MeV$\cite{page} due to the effective threshold being that of $DD^{**}$,
and so the feeding of $\psi(3685),\psi$ with significant branching
ratios leading to an effective probability $0(10^{-5}-10^{-4})$ may be
possible when all states and decay chains are summed. The upper end of this
range would correspond to this being the dominant contribution to $\psi$
production at the Tevatron.

\section{Narrow Charmonium}

In potential models of charmonium, additional narrow states are expected
above the $\psi(3685)$. Most probable are $2^{-+},2^{--}$
predicted at 3.81 to 3.82GeV\cite{godfrey,Quigg} and the possibility that
$2^3P_{1,2}(3.9-4.0GeV)$ have suppressed hadronic widths due to quantum
numbers or nodes in form factors manifested in decays near threshold
\cite{ley}.We consider these in turn.

An estimate of the production rate for the $2^{-+}$ follows by assuming
this to be dominated by the $Q\bar{Q}g$ component of the wavefunction.
In this case we take the result in eq(10) and weigh it by $O(v^2) \approx
\frac{1}{4}$, leading to probability

\begin{equation}
10^{-4} \leq P(2^{-+}) \leq 3 \times 10^{-4}.
\end{equation}
We expect that production of the $2^{--}$ will be much
reduced as there is no obvious octet enhancement available, and we shall not
consider this state further here.

The total hadronic width of $2^{-+}$ is expected to be in the range
$50keV - 800keV$. Our argument is as follows.

By analogy with ref\cite{Bra2} we separate the decay into singlet and
octet pieces

\begin{equation}
\Gamma(^1D_2) = H_1^d \times \frac{4\pi \alpha_s^2}{3} + H_8 \times
\frac{\pi n_f \alpha_s^2}{3}
\end{equation}
where\cite{novikov}

\begin{equation}
H_1^d = \frac{R_d''(0)^2}{2\pi m_Q^6} \approx 200keV
\end{equation}
leading to a width estimate, in the absence of colour octet contribution, of
$O(50keV)$. For the octet contribution, $H_8$, we take two values: one is the
 same as used for the $2^{++}$ (eq 5; $H_8 = 3.3 \pm 0.7MeV$)
 and the other will be to reduce this by
$O(v^2) \approx \frac{1}{4}$ for the reasons outlined prior to eq(9).
Thus for the octet component alone the width would range between
$120keV \leq \Gamma_8 \leq 750keV$. Note that the small (large) widths
correlate with small (large) production probabilities; hence the ratio of
these quantities is less uncertain (and is the relevant quantity that will
enter in our estimate of the $\psi$ yield).

Combining the singlet and octet contributions leads to a width estimate

\begin{equation}
170keV \leq \Gamma (2^{-+}) \leq 800keV
\end{equation}
(note that this also suggests a hybrid $2^{-+}$ annihilation width of
$O(1MeV)$)

To estimate the branching ratios into $\psi(3685), \psi$ and cascades
via $^1D_2 \rightarrow \gamma + ^1P_1; ^1P_1 \rightarrow \psi + \ldots$
we use refs\cite{novikov,yan2}, rescaled for a $^1D_2 \approx 3.81GeV$.

The $^1D_2 \rightarrow \psi(3685) + \gamma$ is suppressed by wavefunction
orthogonality unless there is $^3D_1$ presence in the $\psi(3685)$. The
rate is

\begin{equation}
\Gamma(^1D_2 \rightarrow \psi(3685) + \gamma) = O(2keV) \times p_d
\end{equation}
where $p_d$ is the $D$-wave probability. This implies a branching ratio
below $1\%$. This could double the $\psi(3685)$ strength but does not
seem helpful for explaining a 30-fold enhancement unless the intrinsic
production rate of the $^1D_2$ is larger than our estimates.

Although this is unlikely to resolve the $\psi(3685)$ anomaly it does
offer the hope of isolating some ``missing" charmonium states in the
Tevatron and $B \rightarrow \psi + \ldots$ data. In this regard,the
cascades to $\psi(3095)$ may be interesting. Ref\cite{novikov}
predicts

\begin{equation}
\Gamma(^1D_2 \rightarrow ^1P_1 + \gamma) = 12keV (\frac{k}{100MeV})^3
\end{equation}
where $k$ is the photon momentum. The resulting branching ratio could
be $50\%$. According to ref\cite{yan2} $B.R.(^1P_1 \rightarrow \psi + \dots)
\approx 1\%$ and so

\begin{equation}
B.R.(^1D_2 \rightarrow \psi + \ldots) \approx (4 \pm 3)10^{-3}
\end{equation}

These estimates suggest that forming mass plots of $\psi + \pi,2\pi,\gamma,
\eta$ etc.
may isolate $^1P_1$ and $^1D_2$ states, and that these will make some
additional contribution to the $\psi$ signal at the Tevatron but are
unlikely alone to explain the full anomaly.

The final candidate is the production and decay of radial charmonium, in
particular $2^3P_{1,2} \rightarrow 2^3S_1 + \gamma$. The $2^3P_2$ is predicted
to lie 200-300MeV above $D\bar{D}$ threshold but can have a reduced width
due to the $D$-wave phase space and also dynamical effects associated with
radial excitations\cite{ley,koko,page1}. The $^3P_1 \rightarrow DD^*$ also
is near threshold and the $S$ and $D$-waves are affected by radial wavefunction
nodes which can conspire to reduce the width and, in consequence, increase
the branching ratios  $2^3P_{1,2} \rightarrow 2^3S_1 + \gamma$.

The production probabilities of $2^3P_J$ are estimated as for
their $1^3P_J$ counterparts (eq5 {\it et seq}). The magnitude of $H_1$
for the radial is similar to the $1P$ states\cite{godfrey}; the magnitude
of the octet contribution is less certain. If it is reduced by radial effects
in proportion to the well known colour singlet $S$-state wavefunctions
there is a partial cancellation with the $H_1$ contribution and hence
an overall reduction relative to the $1P$ of two($^3P_1$) to three($^3P_2$).
This would imply a probability of $10^{-4}$ for each of these states in
gluon jets, which is still a factor of 100 greater than the {\it direct}
fragmentation into $\psi(3685)$. An order of magnitude increase in the
$\psi(3685)$ signal relative to theory would therefore follow if the
$BR(2^3P_{1,2} \rightarrow 2^3S_1 + \gamma) \approx 10\%$.
(A similar conclusion has been made in ref\cite{cho})

We can estimate the absolute values of the radiative widths by rescaling the
known values\cite{PDG94} for the bottomonium system. Allowing a factor of two
uncertainty for wavefunction differences between
 $c\bar{c}, b\bar{b}$\cite{godfrey} we anticipate $\Gamma(2^3P_{2c}
\rightarrow 2^3S_{1c} \gamma) \approx 80-160keV$ and a similar magnitude for
the
$2^3P_1$ case. Thus these states are likely to be serious candidates for
the $\psi(3685)$ enhancement only if their total widths are $\approx 1-2MeV$.

In flux tube\cite{koko} or $^3P_0$ models\cite{ley} the decays of radially
excited states can be artificially suppressed if the momenta of the
produced $DD,DD^*$ coincide with nodes in the form factors. The $D$-wave
decay of $^3P_2$ and the $S,D$-wave decays of $^3P_1 \rightarrow DD^*$ have
widths that are rather sensitive to the parameters and masses. If we choose
parameters by fitting the hadronic decays of $\psi(3772)$, $^3D_1$ and other
candidate excited $\psi$ states\cite{page1} the resulting $D$-wave
contributions for the $^3P_J$ decays are in the range $2-5MeV$. The $S$-wave
contributions are highly sensitive to the positions of the wavefunction nodes
and hence, in particular, the masses of the states. If the mass is taken from
ref\cite{godfrey}, $^3P_1=3.95GeV$ then $\Gamma_S \approx 2.5MeV$;
if the mass is
slightly higher (3.96GeV) the $S$-wave hits a node and the width is killed.

Thus model calculations of the hadronic widths suggest that
$^3P_2$ has width $\approx 2-5MeV$, and that the $^3P_1$ width may
be similar though this latter result is highly sensitive to the choice of
parameters.

Our conclusion is that $^1D_2,^3D_2$ charmonium states may be present in the
CDF data and detectable but are unlikely to explain the $\psi(3685)$
enhancement. The radially excited states $2^3P_{1,2}$ may have
suppressed widths enabling them to make a measurable contribution, though
unlikely on their own to explain the full signal.If the colour octet production
mechanism is important, as refs\cite{Bra1,Bra2,Bra} have argued, and if
hybrid states containing dominant gluonic excitations exist, then these
states are likely to make a significant contribution to the $\psi$
and $\psi'$ signals at the Tevatron, and may also be present in the
$B \rightarrow \psi + \ldots$. While our estimates are at this stage
only approximate, they do suggest that these various states may be
present at the right order of magnitude to warrant a serious study.
Our results suggest that a significant percentage of $\psi,\psi'$
production at the Tevatron, in $B$-decay, or in gluon jets at LEP
and HERA, could come from higher mass metastable charmonium and that
invariant mass plots including the $\psi,\psi(3685)$ may reveal new
charmonium states.

\section{Acknowledgements}

I am indebted to E.Braaten, J.Dowell,K.Ellis, P.Kabir, P,Litchfield,
M.Mangano, P.Page and A.Pilaftsis for discussions and comments.


\newpage
\begin{thebibliography}{30}




\bibitem{CDF} M.Mangano, CDF Collaboration, presented at XXIX Rencontres de
Moriond on QCD and High energy Hadronic Interactions, Meribel, March 1994;
and at 27th Internatioanl Conference on High Energy Physics, Glasgow,
July (1994)


\bibitem{godfrey} S.Godfrey and N.Isgur Phys Rev {\bf D32} (1985) 189


\bibitem{Quigg} W.Kwong, J.L.Rosner and C.Quigg, Ann Rev Nuclear and Particle
Science, {\bf 37} (1987) 325

H.Harari, Physics Letters {\bf 64B} (1976) 469


\bibitem{ley} A.LeYaouanc, L.Oliver, O.Pene and J.C.Raynal Physics Letters
{\bf 71B} (1977) 397

\bibitem{koko} R.Kokoski and N.Isgur, Phys Rev {\bf D35} (1987) 907

\bibitem{page} F.E.Close and P.Page, RAL Report (in preparation)

\bibitem{has} P.Hasenfratz, R.Horgan, J Kuti and J.Richard, Phys Lett {\bf 95B}
(1980)299

\bibitem{BCD} T.Barnes and F.E.Close, Phys Letters {\bf 116B} (1982)365;

M.Chanowitz and S.Sharpe Nucl Phys {\bf B222} (1983)211

T.Barnes, F.E.Close and F. de Viron, Nucl Phys {\bf B224}(1983)241.

M.Flensberg, C.Petersen and L.Skold, Z.Phys {\bf C22}(1984) 293


\bibitem{IP} N.Isgur and J.Paton,Phys Rev {\bf D31} (1985) 2910;
N.Isgur, R.Kokoski and J.Paton, Phys Rev Lett {\bf 54} (1985) 869


\bibitem{mich} C.Michael and S.Perantonis, Nucl Phys {\bf B347} (1990) 854

\bibitem{BCS} T.Barnes, F.E.Close and E.Swanson, RAL Report (in preparation)

\bibitem{slear} F.E.Close, Physics at SuperLEAR, Institute of Physics
Conference Series Number 124 (C.Amsler and D Urner eds) (1991) p63

\bibitem{Lipkin}
F.E.Close and H.J.Lipkin, Phys Lett {\bf 196B} (1987) 245

\bibitem{Iddir} F.Iddir et al Phys Lett {\bf B205} (1988) 564;
A.Le Yaouanc et al Z.Phys {\bf C28} (1985) 309

\bibitem{page1} P.Page, unpublished

\bibitem{Bra1} G.T.Bodwin, E.Braaten, T C Yuan and G.P.Lepage, Phys Rev
{\bf D46},(1992),R3703


\bibitem{Bra2} G.T.Bodwin, E.Braaten and G.P.Lepage, {\bf D46},(1992) 1914

\bibitem{Bra} E.Braaten and T.C.Yuan, Phys Rev Letters {\bf 71} (1993) 1673;
Fermilab-PUB-94/040-T (hep-ph/9403401)


\bibitem{Li} F.E.Close and Z.P.Li, Phys Rev Letters {\bf 66} (1991) 3109;
Z.Phys {\bf C54} (1992) 147




\bibitem{Roy} D.P.Roy and K.Sridhar CERN-TH-7329/94

\bibitem{grec} M.Cacciari and M.Greco, Frascati preprint LNF-94/024


\bibitem{bra3} E.Braaten, M.A.Doncheski, S.Fleming and M.Mangano, Fermilab
report FERMILAB-PUB-94/135-T

\bibitem{UA1} UA1 Collaboration, C.Albajar et al.
Physics Letters {\bf B256} (1991) 112


\bibitem{PDG94} Particle Data Group (1994) Advance Summary sheet, presented
at 27th International Conf on HEPP, Glasgow, August(1994)

\bibitem{Narison} S.Narison, ``QCD Spectral Sum Rules", Lecture Notes in
Physics Vol26,p375 (World Scientific,1989)

\bibitem{yan} Y.P. Kuang and T.M.Yan, Phys Rev {\bf D41} (1990) 155


\bibitem{novikov} V.A.Novikov et al Physics Reports {\bf 41C} (1978) 1

\bibitem{yan2} Y.P.Kuang, S.F.Tuan and T.M.Yan, Phys Rev {\bf D37} (1988) 1210


\bibitem{cho} P.Cho, M.B.Wise and S.P.Trivedi, Caltech report CALT-68-1943,
hep-ph/9408352

\end{thebibliography}
\end{document}